\documentclass[letterpaper]{article} 
\usepackage{aaai24}  
\usepackage{times}  
\usepackage{helvet}  
\usepackage{courier}  
\usepackage[hyphens]{url}  
\usepackage{graphicx} 
\usepackage{natbib}  
\usepackage{caption} 
\usepackage{algorithm}
\usepackage{algorithmic}
\usepackage{enumitem}
\usepackage{amsmath}
\usepackage{amsfonts} 
\usepackage{booktabs}
\usepackage{multirow}
\usepackage{bm}
\usepackage{newfloat}
\usepackage{listings}
\DeclareCaptionStyle{ruled}{labelfont=normalfont,labelsep=colon,strut=off} 
\floatstyle{ruled}
\newfloat{listing}{tb}{lst}{}
\floatname{listing}{Listing}
\title{Audio Generation with Multiple Conditional Diffusion Model}
\author {
    Zhifang Guo$^{1,2}$,
    Jianguo Mao$^{1,2}$ \\
    Rui Tao$^{2,3}$,
    Long Yan$^{3}$,
    Kazushige Ouchi$^{3}$,
    Hong Liu$^{1}$,
    Xiangdong Wang$^{1}$
    \thanks{Corresponding author.}
}
\affiliations {
    $^1$Bejing Key Laboratory of Mobile Computing and Pervasive Device,\\
    Institute of Computing Technology, Chinese Academy of Sciences, Beijing, China\\
    $^2$University of Chinese Academy of Sciences, Beijing, China \\
    $^3$Toshiba China R\&D Center, Beijing, China\\
    guozhifang21s@ict.ac.cn, mao@maojianguo.cn \\
    \{taorui, yanlong\}@toshiba.com.cn, kazushige.ouchi@toshiba.co.jp,
    \{hliu, xdwang\}@ict.ac.cn
}

\usepackage{bibentry}

\begin{document}

\maketitle

\begin{abstract}

Text-based audio generation models have limitations as they cannot encompass all the information in audio, leading to restricted controllability when relying solely on text. To address this issue, we propose a novel model that enhances the controllability of existing pre-trained text-to-audio models by incorporating additional conditions including content (timestamp) and style (pitch contour and energy contour) as supplements to the text. This approach achieves fine-grained control over the temporal order, pitch, and energy of generated audio. To preserve the diversity of generation, we employ a trainable control condition encoder that is enhanced by a large language model and a trainable Fusion-Net to encode and fuse the additional conditions while keeping the weights of the pre-trained text-to-audio model frozen. Due to the lack of suitable datasets and evaluation metrics, we consolidate existing datasets into a new dataset comprising the audio and corresponding conditions and use a series of evaluation metrics to evaluate the controllability performance. Experimental results demonstrate that our model successfully achieves fine-grained control to accomplish controllable audio generation. Audio samples and our dataset are publicly available\footnote{\url{https://conditionaudiogen.github.io/conditionaudiogen/}}.

\end{abstract}

\section{Introduction}

\begin{figure}[!t]
    \centering
    \includegraphics[width=0.47\textwidth]{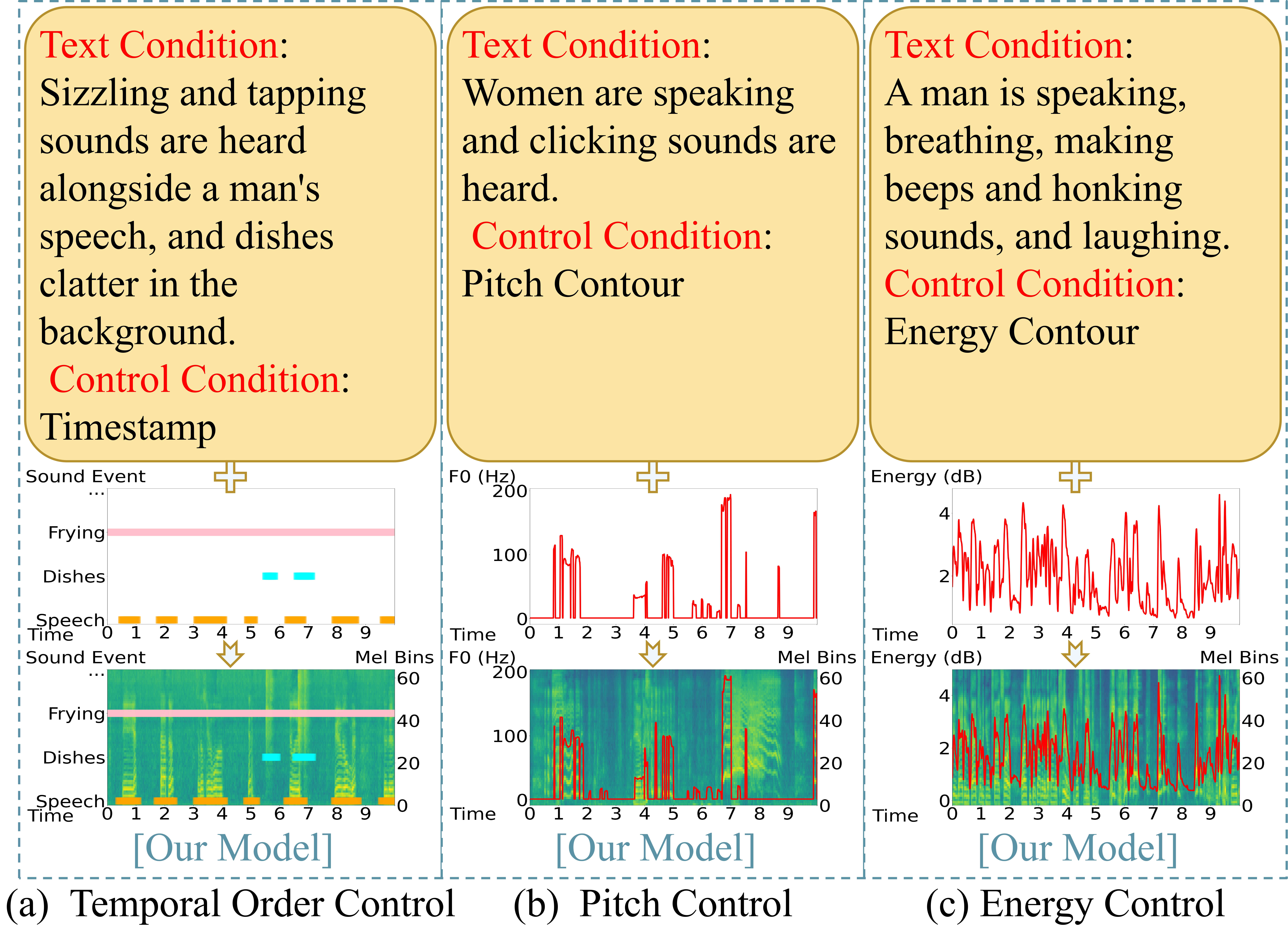}
    \caption{Our model supports fine-grained control by enabling the existing TTA models to accept control conditions. We show temporal order control in Subfigure (a), pitch control in Subfigure (b), and energy control in Subfigure (c).}
    \label{show}
\end{figure}

In recent years, significant progress has been made in generation models, with enhanced naturalness observed across diverse domains such as image~\cite{ramesh2022dalle2,chen2020igpt,nichol2022glide} and text~\cite{radford2020gpt3,schick2023peer,kong2020panns}. This is due to the more stable learning objectives of the models and the availability of extensive paired data. In addition to these fields, audio generation has also attracted significant attention, with strong potential for applications in video creation, virtual reality, and interactive systems.

Audio generation begins with label-to-audio~\cite{fullband2023pascual,csg2021liu}, but the diversity of generated audio is limited by the classes of labels. Considering that text is a more natural way to express the need of audio, recent research on audio generation primarily focuses on using captions (referred to as \textit{text condition}) to generate semantically coherent audio~\cite{yang2022diffsound,kreuk2022audiogen,liu2023audioldm,huang2023makeanaudio,ghosal2023tango}. However, in other domains, many generation models have been proposed to explore the use of other conditions (referred to as  \textit{control condition}), such as layout~\cite{li2021collagingcg} and keypoint~\cite{he2023latentkeypointgan}, to control the generation process, even employing multiple conditions simultaneously~\cite{li2023gligen}. These control conditions serve as a guidance for the generation process to approximate specific distributions. However, within the audio domain, there is little research on audio generation with multiple conditions that are necessary for enriching the expression of the text condition. For instance, when generating audio for video or virtual reality, it is important to specify the temporal locations of sound events and precisely control the pitch and energy to enhance users' immersion in the scene. Obviously, it is difficult to precisely convey the intended fine-grained control over the audio through language alone. Therefore, it is reasonable to use other control conditions as a supplement to the text condition to overcome the limitations of language expression and facilitate the understanding of users' intention by the model. As shown in Figure~\ref{show}, our focus is on three crucial control conditions: timestamp, pitch contour, and energy contour. These conditions are used to control the temporal order of sound events, as well as their pitch and energy, respectively. Temporal order can control the content of the audio, while pitch and energy can control its style. All of these factors are important characteristics for audio. 

Inspired by recent advances in the grounded text-to-image model GLIGEN~\cite{li2023gligen}, we present a new model that expands the capabilities of existing text-to-audio (TTA) models by incorporating additional control conditions while preserving the original text condition for fine-grained control.  Our approach first employs a shared control condition encoder to perform encoding for all control conditions. We leverage the success of large language models (LLMs) across various domains~\cite{chen2021pix2seq,radford2021learning,ruan2023accommodatingam} and utilize their semantic understanding capabilities to obtain distinguishable semantic representations of sound event classes. To ensure continuity of sound events, we adopt frame-level semantic representations.  To preserve the diversity of generation while enabling the pre-trained model to support control conditions, we freeze the weights of the pre-trained model ~\cite{zhang2023adding,li2023gligen} and train a Fusion-Net to integrate control condition information into the audio generation process. Additionally, we enable one model to support multiple control conditions, allowing parameter sharing and enhancing feature representation, resulting in benefits for each type of control conditions and parameter efficiency.

As the first exploration of audio generation with multiple conditions, there currently lacks datasets and evaluation metrics designed for this task. Therefore, we design a dataset and evaluation metrics as a benchmark for future research: (1) Dataset: We integrate existing datasets to create a new dataset for this task, which contains audio, corresponding text, and control conditions. (2) Evaluation metrics: We employ sound event detection (SED) system detection results to measure the performance of temporal order control. Moments are used to assess the pitch distribution of generated audio, and mean absolute error (MAE) is employed to calculate the similarity between the energy of generated audio and the corresponding reference value. Additionally, we ask experts to listen to the location of sound events as a subjective assessment to complement the results of SED system.

The contributions are summarized as follows:
\begin{itemize}
\item We introduce a new task that generates audio guided by both text and control conditions, enabling fine-grained customization of the audio with timestamp, pitch contour, and energy contour.
\item We integrate the existing datasets to create a new dataset comprising the audio and corresponding conditions and use a series of evaluation metrics to assess performance, which can serve as a benchmark for future work.
\item We propose an audio generation model based on existing pre-trained TTA models, which accepts not only text as a condition but also incorporates other control conditions to achieve finer-grained and more precise control on audio generation. Experimental results demonstrate the effectiveness of our model in generating audio with greater control. 
\end{itemize}

\begin{figure*}[!t]
    \centering  
    \includegraphics[width=0.80\textwidth]{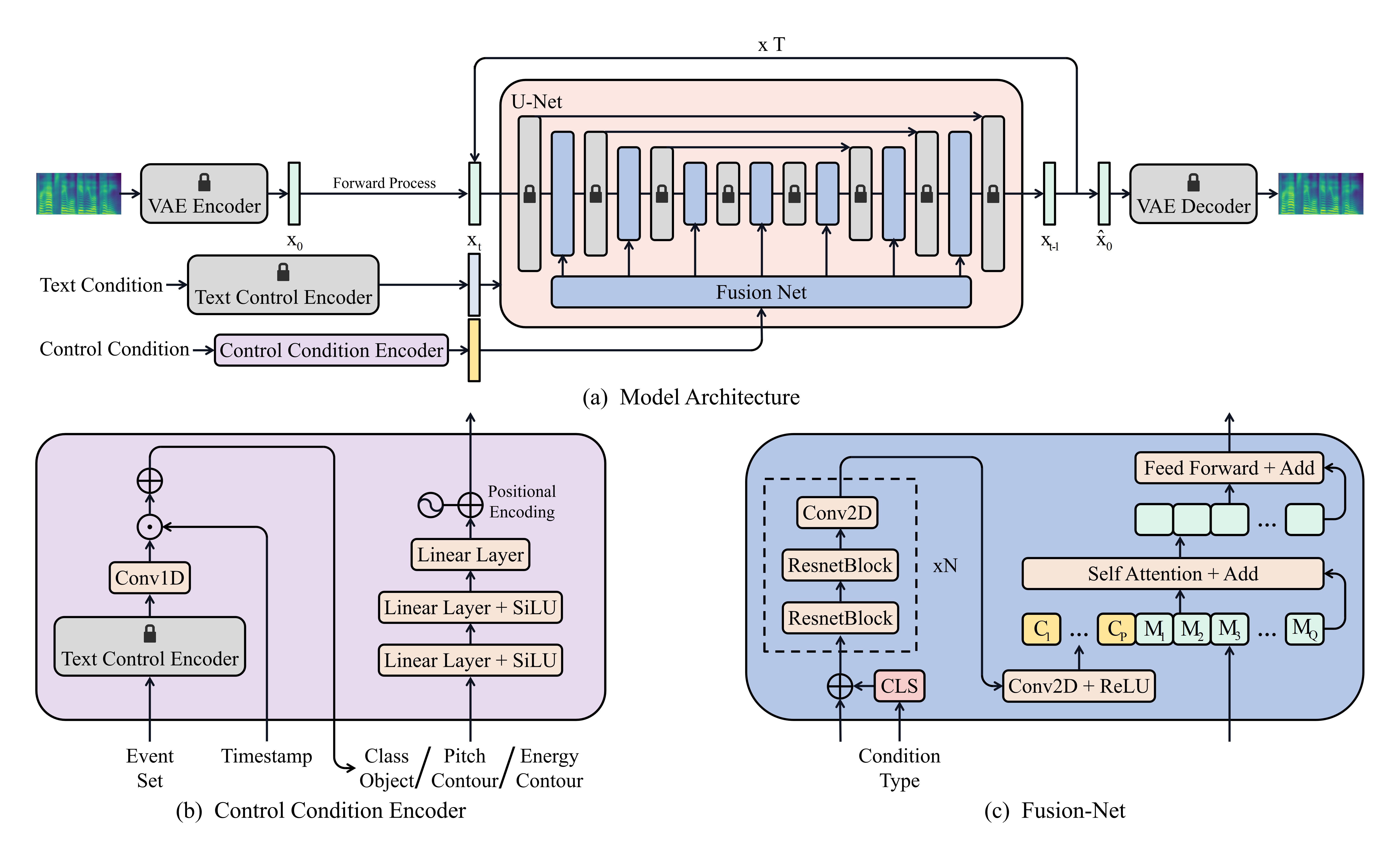}
    \caption{The overview of our model. Subfigure (a) is the architecture of our model that synthesizes audio guided by the text and control conditions. Subfigure (b) shows the control condition encoder that takes the control conditions including timestamp, pitch contour, and energy contour as input and extract control embedding to guide the audio generation. Subfigure (c) demonstrates the Fusion-Net between each layer of the U-Net, enabling the fusion of control embedding.}
    \label{model}
\end{figure*}

\section{Related Work}

\subsection{Text-Guided Audio Generation}

Diffusion models~\cite{jonathan2020ddpm} have demonstrated remarkable success in generation models across various fields, including image~\cite{chen2020wavegrad}, speech~\cite{kong2021diffwave}, and video~\cite{singer2022makeavideo}. However, the iterative processing of high-dimensional features in diffusion models can lead to slow inference speed. To address this challenge, researchers reduce the dimensions of features during the iteration process and most audio generation models are built on this foundation: first, an encoder is employed to convert audio features into latent representation. Subsequently, a diffusion model operates within this latent space, and the predicted outcomes are rebuilt back to audio features with a decoder. Notably, various improvements have been built upon this foundation: DiffSound~\cite{yang2022diffsound} introduces VQ-VAE~\cite{van2017vqvae} in a discrete diffusion model to process mel-spectrograms. Instead of training over mel-spectrograms, AudioGen~\cite{kreuk2022audiogen} learns directly from the raw waveform and utilizes classifier free guidance~\cite{jonathan2022classfree} under an autoregressive structure for audio generation. More recently, Make-an-Audio~\cite{huang2023makeanaudio} and AudioLDM~\cite{liu2023audioldm} have conducted an exploration of audio generation within a continuous space. In addition, to enhance training efficiency, Tango~\cite{ghosal2023tango} utilizes the language under
ing capabilities from a frozen pre-trained LLM to train a diffusion model on a limited training dataset. However, the current conditions mainly involve text, which poses challenges to conveying additional information by natural language. Therefore, we enhance the control capabilities of the text-guided audio generation model by enabling it to be guided by additional conditions including content (timestamp) and style (pitch contour and energy contour), which is simple for users to create the audio that meets their requirements.

\subsection{Conditional Generation Model}

With the increasing availability of diverse paired data, there has been rapid progress in the development of generation models conditioned by various types of information, such as text~\cite{ramesh2022dalle2,liu2021diffsinger}, image~\cite{huang2023makeanaudio,li2023blip2}, audio~\cite{zelaszczyk2021audiotoimage,wang2023one}, and
layout~\cite{jahn2021highresolutioncs,li2021lama}, but they rely solely on one condition, limiting the information transmitted. Therefore, it is crucial to explore how to leverage the existing knowledge within these generation models to extend to new conditions. GLIGEN~\cite{li2023gligen} revolutionize text-to-image models by incorporating grounding conditions to control the generation of image and preserving the original text condition. We want to explore the possibilities of extending conditions in the audio domain. In addition, we also want to excavate the relationships between these different conditions and their effects on the audio generation.

\section{Method}

In this section, we begin with presenting the architecture of our model. Subsequently, we introduce the method and specific details for handling control conditions. Finally, we illustrate the method employed to augment the existing TTA model with control conditions.

\subsection{Overview}

Our model consists of a text condition encoder, a control condition encoder, a conditional latent diffusion model (LDM), a variational auto-encoder (VAE)~\cite{kingma2013vae}, and a Fusion-Net. The text and control condition encoders extract the corresponding embeddings, respectively. Under the guidance of the text and control embeddings, the noised mel embedding (for training) from the VAE encoder or standard Gaussian noise (for testing) is used to construct latent representation , which is rebuilt back to the mel-spectrogram by the VAE decoder. And we use a vocoder to generate the final audio. In order to preserve the generation ability, we freeze the VAE, text condition encoder, and conditional latent diffusion model and use the weights from a pre-trained TTA model, Tango~\cite{ghosal2023tango}. The overview of our model is shown in Figure~\ref{model} (a).

\subsubsection{Text Condition Encoder} As the instruction-tuned LLM makes a significant improvement in the performance of the TTA model~\cite{ghosal2023tango}, we employ the pre-trained LLM, \textit{FLAN-T5-LARGE}~\cite{chung2022siflm}, as the text condition encoder to generate text embedding with rich semantic information. And the parameters of the text condition encoder are frozen during the training stage to keep the original semantic representation ability.

\subsubsection{Control Condition Encoder} As shown in Figure~\ref{model} (b), the control condition encoder takes the control condition as input, including timestamp, pitch contour, and energy contour. For the temporal order control, we leverage the language knowledge derived from \textit{FLAN-T5-LARGE} to generate the semantic representation of the event classes and convert it with timestamp into the class object. Then the class object, pitch contour, and energy contour are processed using the same multi-layer perceptron and position embedding to generate the control embeddings of the same dimensions, which will be discussed in \textit{Subsection Control Condition Preprocessing}.

\subsubsection{Conditional Latent Diffusion Model} Our model can be seen as a LDM that relies on the conditions including the noised mel embedding $\mathbf{x}_{t}$, the text embedding $\mathbf{c}_{\text{text}}$, and the control embedding $\mathbf{c}_{\text{control}}$. The mel embedding $\mathbf{x}_{t}$ is obtained through the diffusion (forward) processing of the initial mel embedding $\mathbf{x}_{0}$ with variance preserving:

\begin{align}
\mathbf{x}_{t} = \sqrt{\overline\alpha_{t}} \mathbf{x}_{0} + (1-\overline\alpha_{t})\epsilon
\end{align}

where $t$ is uniformly sampled from time steps $\{1, \cdots, T\}$, $\epsilon$ is standard Gaussian noise, and $\overline\alpha_{t} = \prod_{s=1}^{t}(1 - \beta_{s})$ is based on the predefined noise schedule $\beta_{s}$. Our aim is to produce denoising (reverse) process on the mel embedding $\mathbf{x}_{t}$ or the standard Gaussian noise $\epsilon$ back to the mel embedding $\mathbf{\hat{x}}_0$. For each step $t$, the LDM training objective solves the denoising problem:

\begin{align}
\mathcal{L}_{\text{LDM}} = \mathbb{E}_{x, \epsilon \sim  \mathcal{N}(\mathbf{0}, \mathbf{I})}
\| \epsilon_{\theta}(\mathbf{x}_{t}, t, \mathbf{c}_{\text{text}}, \mathbf{c}_{\text{control}}) -   \epsilon \|^2_2,
\end{align}

where $\epsilon_{\theta}$ is the $(t, \mathbf{c}_{\text{text}}, \mathbf{c}_{\text{control}})$-conditioned noise estimation. The LDM architecture primarily relies on a U-Net~\cite{yang2022diffsound} with a cross attention mechanism~\cite{vaswani2017attention} comprising a series of ResNet~\cite{he2016resnet} and Transformer blocks~\cite{vaswani2017attention}. Similarly to the standard diffusion model, it takes the noised mel embedding $\mathbf{x}_{t}$ and time step $t$ as input conditioned by the text embedding $\mathbf{c}_{\text{text}}$ to predict the noise estimation $\epsilon_{\theta}$. Within each Transformer block, a cross attention layer is incorporated to fuse the time step $t$ and text embedding $\mathbf{c}_{\text{text}}$. In order to preserve the generation capability of the original TTA model, we freeze the parameters of each layer of the U-Net.

\subsubsection{Fusion-Net} In order to incorporate the control embedding $\mathbf{c}_{\text{control}}$ into the LDM, we introduce a trainable Fusion-Net between each layer of the frozen U-Net, which will be discussed in \textit{Subsection Control Condition Fusion}.

\subsubsection{Variational Auto-Encoder} The VAE~\cite{kingma2013vae} consists of an encoder and a decoder that compresses the mel-spectogram into the mel embedding $\mathbf{x}_{0}$ and reconstructs the mel-spectogram from the mel embedding $\hat{x}_0$. The VAE is constructed using ResUNet blocks~\cite{kong2021decouplingma} and are trained using a combination of ELBO~\cite{kingma2013vae} and adversarial loss~\cite{isola2017itiadversarial}. We use the VAE checkpoint from AudioLDM~\cite{liu2023audioldm} and freeze its parameters.

\subsection{Control Condition Preprocessing}
\label{control condition preprocessing}

In audio generation, besides the text condition, there are various other conditions that can control the generation process. Our control conditions include timestamp, pitch contour, and energy contour and we demonstrate how to extract and process these control conditions in this section.

\subsubsection{Timestamp} Previous layout-to-image models~\cite{li2023gligen,jahn2021highresolutioncs,li2021lama} commonly employ the bounding box labeled with classes to control the positions of objects in generated image, which map these inputs into a series of fixed-length tokens and every token contains not only category information but also top left and bottom right coordinates. However, when dealing with audio, more accurate representation of timestamp is necessary, as audio generation requires more continuous sound events.

As shown in Figure~\ref{model} (b), the event set is a collection of sound event classes with the number of $D$ and the frame-level timestamp $\mathbf{i}_{\text{timestamp}} \in \mathbb{R}^{D \times L}$ is a matrix and $L$ represents the number of frames, within which 1 indicates the presence of a sound event and 0 indicates the absence of a sound event. To distinguish sound events from a semantic perspective, we firstly employ a frozen \textit{FLAN-T5-LARGE}~\cite{ghosal2023tango,chung2022siflm}, which is identical to the text condition encoder, to convert the sound event classes into the semantic representation. To maintain the independence of each sound event, we utilize a trainable $1 \times 1$ convolutional layer to convert the output of \textit{FLAN-T5-LARGE} into a label embedding $\mathbf{i}_{\text{label}} \in \mathbb{R}^{D \times H}$ and $H$ refers to the hidden size of the label embedding. Finally, we multiple the label embedding $\mathbf{i}_{\text{label}}$ with with the corresponding position of the frame-level timestamp and sum the dimensions of sound event classes to obtain the frame-level semantic representation class object $\mathbf{i}_{\text{object}} \in \mathbb{R}^{L \times H}$:

\begin{align}
\mathbf{i}_{\text{object}} = \sum_{D}{\mathbf{i}_{\text{label}} \odot \mathbf{i}_{\text{timestamp}}}
\end{align}

\subsubsection{Pitch Contour} Pitch is a crucial characteristic of sound effects, music, and speech, but current TTA models primarily rely on the text for coarse-grained control~\cite{liu2023audioldm} and we aim to achieve fine-grained control by incorporating the pitch contour $\mathbf{i}_{\text{pitch}} \in \mathbb{R}^{L \times H}$. In the domain of text-to-speech, pitch contour exhibits considerable variations and the distribution of pitch poses challenges for models to comprehend~\cite{ren2021fastspeech2}. Therefore, we employ the continuous wavelet transform to decompose the continuous pitch contour and quantize each frame of it into 256 possible values on a logarithmic scale~\cite{wuni2013wavelets,ren2021fastspeech2}, which serves as the control condition and is converted into the control embedding by the control condition encoder.

\subsubsection{Energy Contour} Energy control is equally important in the field of text-to-speech and we incorporate energy contour $\mathbf{i}_{\text{energy}} \in \mathbb{R}^{L \times H}$ as a control condition during audio generation. Drawing inspiration from the method outlined in Fastspeech 2~\cite{ren2021fastspeech2}, we calculate the energy of each short-time Fourier transform frame by computing the L2-norm of its amplitude and quantizing on a logarithmic scale, which is encoded using the control condition encoder.

In order to achieve one model supporting multiple control conditions (timestamp, pitch contour, and energy contour), we also apply the operations above to standardize the control conditions to the same dimensions. Subsequently, we utilize a shared multi-layer perceptron to encode them, employing the condition type as a prompt to distinguish different control conditions within the Fusion-Net.

\subsection{Control Condition Fusion}
\label{control condition fusion}

The existing TTA models have been pre-trained on large scale datasets comprising text-audio pairs, enabling them to generate audio based on diverse and complex text. Our objective is to maintain the high-quality audio generation capability of these large models, which have been trained on billions of audio samples, while also enhancing their capability to support additional conditions. Similarly to the practice in the image field~\cite{zhang2023adding,li2023gligen}, we freeze the original weights, and gradually adapt the model by tuning the Fusion-Net.

As shown in Figure~\ref{model} (c), the Fusion-Net takes the control embedding as input added with the condition type embedding $CLS$. We adjust the dimension of the control embedding to match that of the mel-spectogram, after a series of structures resembling the VAE encoder, which includes ResnetBlocks and convolutional layers~\cite{kingma2013vae}. The 2-D convlutional layers with different kernel sizes and strides are applied to generate control condition tokens ($C_1, ..., C_P$) of different dimensions. Between each layer of the fixed U-Net, we concatenate the control condition tokens ($C_1, ..., C_P$) with the mel tokens ($M_Q, ..., M_Q$) obtained from the previous layer in the U-Net. Subsequently, we employ a self-attention and only select the output tokens corresponding to the mel tokens for the feed forward block~\cite{li2023gligen}. 

\subsection{Classifier-Free Guidance}

During the process of denoising (reverse) process, we incorporate the classifier-free guidance~\cite{jonathan2022classfree} of text embedding $\mathbf{c}_{\text{text}}$ and control embedding $\mathbf{c}_{\text{control}}$ to control the degree of guidance:

\begin{align}
\hat{\epsilon}_{\theta} = \omega\epsilon_{\theta}(\mathbf{x}_{t}, t, \mathbf{c}_{\text{text}}, \mathbf{c}_{\text{control}}) + (1 - \omega)\epsilon_{\theta}(\mathbf{x}_{t}, t), 
\end{align}

where $\omega$ refers to the guidance scale. To reduce the model's reliance on guidance, we randomly exclude the guidance for 10\% of the training samples. 

\section{Dataset}

Given that there are no TTA datasets with both text and control conditions, we integrate the existing datasets to a new dataset called \textit{AudioCondition} that consists of the audio along with corresponding conditions. We use audio in AudiosetStrong~\cite{hershey2021audiosetstrong} which contains audio from Audioset~\cite{gemmeke2017audioset}. As for the text condition, we obtain from WavCaps~\cite{mei2023wavecaps}, a caption dataset based on ChatGPT and processed through a three-stage pipeline to filter noisy data and produce high-quality text, which has designed the text condition for the audio in AudiosetStrong. For the timestamp, we obtain it from AudiosetStrong that annotates audio with frame-level timestamp for 456 sound events. For the pitch contour and energy contour, we extract the values of them from audio with signal processing tools~\cite{world2016morise,librosa2015masanori}. We employ sound event detection (SED) systems to provide evaluation metrics, but the number of sound events supported by the current SED~\cite{janek2021sased, janek2020fbsedt, janek2022pssed} systems is limited. Therefore, we only select audio that includes sound events~\footnote{Alarm bell ringing, blender, cat, dishes, dog, electric shaver toothbrush, frying, running water, speech, and vacuum cleaner.} 
supported by SED systems for test set, but our model supports for controlling all sound event classes. We randomly split AudioCondition into three sets: 89557 samples for training, 1398 samples for validation, and 1110 samples for testing, which are publicly available\footnote{\url{https://conditionaudiogen.github.io/conditionaudiogen/}}.

\section{Evaluation Metrics} 

In this section, we introduce evaluation metrics for temporal order, pitch, and energy control: (1) Temporal order control metrics: inspired by the practice in the image field~\cite{li2023gligen,jahn2021highresolutioncs} which evaluates the location of generated object using a object detection model, we employ a SED system to provide event-based measures (Eb) and clip-level macro F1 score (At) to assess the temporal order control capability~\cite{mesaros2016eventf1}. These metrics evaluate the presence of sound events in the generated audio, as well as the onsets and offsets using the first-place SED system in DCASE 2022 Task 4, PB-SED~\cite{janek2022pssed,janek2020fbsedt,janek2021sased}\footnote{\url{https://github.com/fgnt/pb_sed/}}, on AudioCondition test set. (2) Pitch control metrics: to compare the distribution of audio pitch, we computed several moments used in the speech field~\cite{ren2021fastspeech2} including standard deviation ($\sigma$), skewness ($\gamma$), and kurtosis ($\kappa$)~\cite{bistra2014diffpitch,niebuhr2019measuringas} as well as the average dynamic time warping distance (DTW)~\cite{muller2007dtw} of the pitch distribution between the ground-truth audio and synthesized audio. (3) Energy control metrics: we compute the MAE between the frame-wise energy extracted from the generated audio and the energy in the ground-truth audio~\cite{ren2021fastspeech2}.

\section{Experiment Setup}

In this section, we present the experiment setup, including the model configuration and baseline models. 

\subsection{Model Configuration}

Our model mainly consists of the frozen module, trainable module, and vocoder. In the frozen module, the structure primarily involves the pre-trained TTA model, Tango~\cite{ghosal2023tango},which does not require training. The trainable module comprises the control condition encoder and Fusion-Net. We utilize HiFi-GAN~\cite{kong2020hifigan} as the vocoder for fairness and consistency. We add more detailed configurations in \textit{Appendix A}.

\subsection{Baseline Models}

\subsubsection{Control Performance} Given the absence of previous work in generating audio guided by both text and control condition or event controlling audio generation solely based on the control condition, we compared our model with Tango~\cite{ghosal2023tango} and AudioLDM~\cite{liu2023audioldm} that are able to control temporal order, pitch, and other attributes at a coarse-grained level with the text condition to demonstrate that our model has fine-grained control capability. In order to demonstrate the superior performance of our model, drawing inspiration from the single-condition generation models in other domains~\cite{jahn2021highresolutioncs,li2023blip2,radford2020gpt3}, we also design a baseline model that only accepts the control condition without the text condition called control-condition-to-audio (referred to \textbf{CCTA}) as baseline.

\subsubsection{Module Effectiveness} Since our primary focus on the control condition encoder lies in timestamp, we design two baseline models using the weights from Tango~\cite{ghosal2023tango} to verify the effectiveness of the control condition encoder: \textbf{Tango-Timestamp} uses the architecture of Tango and is fine-tuned on AudiosetStrong with the text condition like \textit{``Music from 4.11 to 10.00 and Squeak from 3.17 to 3.50.''} to control the location of sound events with only the text condition. \textbf{Ours-Box} replaces our model's control condition encoder with GLIGEN's bounding box process~\cite{li2023gligen}, which is fine-tuned on AudioCondition with the text and bounding box conditions. We also design a setting \textbf{Ours-Add} to verify the effectiveness of the Fusion-Net, where we eliminate the Fusion-Net and integrate the control embedding into the mel embedding $\mathbf{x}_{t}$.

\section{Result}

In this section, we evaluate the performance of our model (referred to \textbf{Ours}) and the baseline models in temporal order, pitch, and energy control. Then we compare our model trained under multiple control conditions (that \textbf{Ours} employ this setting by default) and single control conditions (referred to \textbf{Ours-Sin}). We also perform an ablation study on the design of the control condition encoder and Fusion-Net. Furthermore, we investigate the advantages of classifier-free guidance scales and inference steps. Finally, we ask experts to provide a subjective evaluation of the temporal order control as a supplement to the SED systems' detection. In \textit{Appendix B}, we conduct further visual analysis to compare the mel-spectrograms generated by different models.

\subsection{Control Performance}

\begin{table}[!h]\footnotesize
    \centering
    \caption{The control performance between Ours and baseline models. ``GT'' stands for the ground-truth recordings.}
    \label{tab:control}
        \setlength{\tabcolsep}{0.56mm}{
        \begin{tabular}{c|cc|cccc|c}
        \toprule
        \multirow{2}{*}{\textbf{Settings}} & \multicolumn{2}{c}{\textbf{Temporal Order}} & \multicolumn{4}{|c}{\textbf{Pitch}} & \multicolumn{1}{|c}{\textbf{Energy}} \\
         & Eb~$\uparrow$ & At~$\uparrow$ & $\sigma$ & $\gamma$ & $\kappa$ & DTW~$\downarrow$ & MAE~$\downarrow$ \\
        \midrule
        \midrule
         GT & 43.37 & 67.53 & 54.21 & 2.02 & 7.31 & $-$ & $-$ \\
         \midrule     
         AudioLDM & 9.75 & 46.55 & 50.96 & 3.17 & 14.31 & 40.87 & 0.669\\
         Tango & 5.20 & 45.11 & 40.38 & 1.23 & 3.24 & 36.11 &  0.553  \\
         CCTA & 14.57 & 18.27 & 50.85 & 1.80 & 5.74 & 11.24 & 0.230 \\
         Ours & \textbf{29.07} & \textbf{47.11} & \textbf{55.68} & \textbf{1.93} & \textbf{6.77} & \textbf{10.79} & \textbf{0.200} \\
        \bottomrule
        \end{tabular}
        }
\end{table}

\subsubsection{Temporal Order Control Performance} According to an Eb and At in Table~\ref{tab:control}, the following observations can be made: (1) The excellent performance achieved by PB-SED on GT demonstrates that PB-SED can effectively serve as an evaluation metric. (2) The low performance of CCTA indicates that the text condition is crucial for controlling audio generation, and the more input conditions there are, the more helpful it is for the model to understand users' intention. (3) Ours achieves an Eb of 29.07 \% and At of 47.11 \%, indicating its capability to generate temporal-order-controlled audio. (4) Ours surpasses the baseline models in terms of both Eb and At, primarily due to its acceptance of control conditions, enabling a fine-grained control to generate audio in a more consistent style and content.

\subsubsection{Pitch Control Performance} We refer to Table~\ref{tab:control} and observe that the $\sigma$, $\gamma$, and $\kappa$ of the audio generated by Ours are closer to that of GT. Additionally, the DTW of Ours is significantly smaller compared to the other settings. These results indicate that Ours exhibits superior performance in terms of pitch control.

\subsubsection{Energy Control Performance} Based on the results presented in the Table~\ref{tab:control}, it is evident that Ours achieves the lowest MAE value, indicating that the energy of Ours is closer to the reference value, which has the best energy control capabilities.

\subsection{Performance Between Multiple and Single Control Conditions}

\begin{table}[!h]\footnotesize
    \centering
    \caption{The control performance between multiple and single control condition.}
    \label{tab:multi}
        \setlength{\tabcolsep}{0.76mm}{
        \begin{tabular}{c|cc|cccc|c}
        \toprule
        \multirow{2}{*}{\textbf{Settings}} & \multicolumn{2}{c}{\textbf{Temporal Order}} & \multicolumn{4}{|c}{\textbf{Pitch}} & \multicolumn{1}{|c}{\textbf{Energy}} \\
         & Eb~$\uparrow$ & At~$\uparrow$ & $\sigma$ & $\gamma$ & $\kappa$ & DTW~$\downarrow$ & MAE~$\downarrow$ \\
        \midrule
        \midrule
         GT & 43.37 & 67.53 & 54.21 & 2.02 & 7.31 & $-$ & $-$ \\
         \midrule
         Ours-Sin & 29.06 & 46.91 & 51.32 & \textbf{1.95} & 6.76 & 11.23 & 0.267 \\
         Ours & \textbf{29.07} & \textbf{47.11} & \textbf{55.68} & 1.93 & \textbf{6.77} & \textbf{10.79} & \textbf{0.200} \\
        \bottomrule
        \end{tabular}
        }
\end{table}

We conducted a comparative analysis on AudioCondition test set between Ours-Sin that is trained and inferred under a single control condition and Ours that is trained under multiple control conditions and inferred under a single control condition. The results depicted in Table~\ref{tab:multi} demonstrate that the performance of Ours slightly outperforms 
Ours-Sin in most evaluation metrics, indicating that training multiple tasks simultaneously enables information sharing and mutual complementation, leading to performance enhancements. Additionally, Ours-Sin consists of three separate models requiring 3177M parameters, while Ours only needs 1076M parameters, which can reduce redundant parameters and enhance the efficiency.

\subsection{Ablation Study}

\begin{table}[!h]\footnotesize
    \centering
    \caption{The control performance of temporal order with different designs for the control condition encoder.}
    \label{tab:ablation control condition encoder}
        \begin{tabular}{c|cccc}
        \toprule
        \multirow{2}{*}{\textbf{Settings}} & \multicolumn{2}{c}{\textbf{Temporal Order}} \\
         & Eb~$\uparrow$ & At~$\uparrow$  \\
        \midrule
        \midrule
         GT & 43.37 & 67.53\\
         \midrule
         Ours-Box & 4.79 & 40.59 \\
         Tango-Timestamp & 9.45 & 46.33 \\
         Ours-Sin & \textbf{29.11} & \textbf{46.91} \\
        \bottomrule
        \end{tabular}
\end{table}

\begin{table*}[!t]
    \centering
    \caption{The control performance with different classifier-free guidance scales and inference steps.}
    \label{tab:classifier-Free guidance and inference steps}
        \begin{tabular}{cc|c|cc|cccc|c}
        \toprule
        \multirow{2}{*}{\textbf{Guidance}} & \multirow{2}{*}{\textbf{Step}} & \multirow{2}{*}{\textbf{Settings}} & \multicolumn{2}{c}{\textbf{Temporal Order}} & \multicolumn{4}{|c}{\textbf{Pitch}} & \multicolumn{1}{|c}{\textbf{Energy}} \\
        & &  & Eb~$\uparrow$ & At~$\uparrow$ & $\sigma$ & $\gamma$ & $\kappa$ & DTW~$\downarrow$ & MAE~$\downarrow$ \\
        \midrule
        \midrule
         $-$ & $-$ & GT & 43.37 & 67.53 & 54.21 & 2.02 & 7.31 & $-$ & $-$ \\
         \midrule
         1 & 200 & \multirow{4}{*}{Ours} & 11.40 & 30.68 & 46.65 & 1.78 & 5.11 & 14.85 & 0.201 \\
         3 & 200 & & 27.90 & 46.72 & 56.73 & 1.84 & 5.85 & 11,08 & 0.208 \\
         5 & 200 & &  \textbf{29.07} & \textbf{47.11} & \textbf{55.68} & \textbf{1.93} & \textbf{6.77} & \textbf{10.79} & 0.200 \\
         10 & 200 & &  25.76 & 45.78 & 51.15 & 1.79 & 5.82 & 11.67 & \textbf{0.176} \\
         \midrule
         5 & 10 & \multirow{4}{*}{Ours} &  21.54 & 39.32 & 57.39 & 1.84 & 5.88 & \textbf{8.90} & 0.242 \\
         5 & 50 & &  30.02 & 50.14 & 55.74 & 1.81 & 5.55 & 10.34 & 0.209 \\
         5 & 100 & & \textbf{31.26} & \textbf{50.27} & \textbf{55.06} & 1.89 & 6.41 & 10.77 & 0.205 \\
         5 & 200 & &  29.07 & 47.11 & 55.68 & \textbf{1.93} & \textbf{6.77} & 10.79 & \textbf{0.200} \\
        \bottomrule
        \end{tabular}
\end{table*}

\subsubsection{Control Condition Encoder Design} The design of the control condition encoder for the timestamp is a significant contribution of our paper and the encoding process for different control conditions exhibits similarities. Consequently, this section is to show the effectiveness of the control condition encoder by comparing the performance of Tango-Timestamp, Ours-Box, and Ours-Sin to control the location of sound events. The results, as depicted in Table~\ref{tab:ablation control condition encoder}, indicate that the baseline models perform similarly to Ours-Sin in terms of At, thereby demonstrating their ability to generate audio with the desired sound events. However, Ours-Sin outperforms the baseline models significantly in terms of Eb, implying that the baseline models struggle to precisely control the fine-grained temporal order of these sound events.

\begin{table}[!h]\footnotesize
    \centering
    \caption{The control performance with different designs for the Fusion-Net.}
    \label{tab:ablation fusion-net}
        \setlength{\tabcolsep}{0.75mm}{
        \begin{tabular}{c|cc|cccc|c}
        \toprule
        \multirow{2}{*}{\textbf{Settings}} & \multicolumn{2}{c}{\textbf{Temporal Order}} & \multicolumn{4}{|c}{\textbf{Pitch}} & \multicolumn{1}{|c}{\textbf{Energy}} \\
         & Eb~$\uparrow$ & At~$\uparrow$ & $\sigma$ & $\gamma$ & $\kappa$ & DTW~$\downarrow$ & MAE~$\downarrow$ \\
        \midrule
        \midrule
         GT & 43.37 & 67.53 & 54.21 & 2.02 & 7.31 & $-$ & $-$ \\
         \midrule
         Ours-Add & 12.40 & 43.28 & 69.13 & 1.87 & 6.43 & 24.72 & 0.606 \\
         Ours & \textbf{29.07} & \textbf{47.11} & \textbf{55.68} & \textbf{1.93} & \textbf{6.77} & \textbf{10.79} & \textbf{0.200} \\
        \bottomrule
        \end{tabular}
        }
\end{table}

\subsubsection{Fusion-Net Design} To assess the efficacy of Fusion-Net, we substitute it with a summation operation and conduct a comparative analysis of three control conditions. As depicted in Table~\ref{tab:ablation fusion-net}, the findings demonstrate that Ours surpasses the Ours-Add across all metrics, indicating that Fusion-Net plays a crucial role in integrating control conditions into the existing TTA model.

\subsection{Effectiveness of Classifier-Free Guidance Scales and Inference Steps}

Ours is based on the weights of Tango whose audio quality is related to the classifier-free guidance scale and inference steps~\cite{ghosal2023tango}. Similarly on AudioCondition test set in Table~\ref{tab:classifier-Free guidance and inference steps}, we have observed that these two parameters also play a significant role in control: (1) Classifier-free guidance scales: in the absence of classifier-free guidance whose scale is set to 1, the performance is poor only when the control condition is the timestamp. By increasing the scale to 5, improved performance is achieved across most evaluation metrics. However, further increment in scale leads to a decline in performance, which may be the diversity brought by larger scales hinders the controllability of the model. (2) Inference steps: temporal order control reaches its optimum at step 100, energy control reaches its optimum at step 200, but pitch control does not have a clearly suitable step that allows all indicators to reach their best values, since pitch could be more difficult to model.

\subsection{Subjective Controllability Evaluation}

\begin{table}[!h]\footnotesize
    \centering
    \caption{The control performance of temporal order with different designs for subjective controllability evaluation.}
    \label{tab:subjective controllability evaluation}
        \begin{tabular}{c|cccc}
        \toprule
        \multirow{2}{*}{\textbf{Settings}} & \multicolumn{2}{c}{\textbf{Temporal Order}} \\
         & Eb~$\uparrow$ & At~$\uparrow$  \\
        \midrule
        \midrule
         GT & 100.00 & 100.00\\
         \midrule
         AudioLDM & 25.45 & 90.00 \\
         Tango & 15.45 & 90.00 \\
         CCTA & 25.75 & 43.33 \\
         Ours-Box & 12.71 & 90.00 \\
         Tango-Timestamp & 22.45 & 90.00 \\
         Ours-Add & 32.71 & 90.00 \\
         Ours-Sin & 76.90 & 96.67 \\
         Ours & \textbf{84.28} & \textbf{96.67} \\
        \bottomrule
        \end{tabular}
\end{table}
 
 Considering that the SED system is employed for this task for the first time, and the number of the sound events supported by the PB-SED system is also limited, we randomly selected 20 audio clips from the test set containing all sound events that are not limited to the 10 sound events detected by PB-SED. And the corresponding generated audio clips are listened by experts to determine the presence of sound events and their onsets and offsets. The results in the Table~\ref{tab:subjective controllability evaluation} show a similar distribution to the results obtained from the SED system above, indicating that the SED system can serve as an evaluation metric for this task. Furthermore, the performance of Ours on other sound events achieves an Eb of 84.28 \% and At of 96.67 \%, demonstrating that Ours can also perform well on all sound events.

\section{Conclusion}

In this work, we aim to investigate the potential for enhancing the current text-to-audio model by improving its control capability. To achieve fine-grained controllable audio generation, we propose a novel model that incorporates not only the text condition but also the control conditions as supplementary information. To facilitate this research, we integrate existing datasets containing audio and corresponding conditions and introduce a series of evaluation metrics serving as a benchmark for future studies in the audio domain. Experimental results demonstrate that our model significantly improves the precision of audio generation control. In the future, we plan to expand our model's control capability to accept a wider range of control conditions, enabling it to process and utilize more diverse supplementary information.


\section{Appendix}

\subsection{A Model Configuration}

\subsubsection{Frozen Module} The frozen modules include the VAE, text condition encoder, and original layers in U-Net. we adhere to the settings in Tango~\cite{ghosal2023tango} and use its weights, which is only trained on the Audiocaps~\cite{kim2019audiocaps}, but the performance surpasses that of many models trained on larger datasets. The VAE accepts audio with 16KHz
frequency and reduce the feature dimension of mel-spectrum by 4 times. The text condition encoder is based on the pre-trained LLM, \textit{FLAN-T5-LARGE}~\cite{chung2022siflm}, which has a total of 780M parameters. The original layers in U-Net are derived from the stable diffusion U-Net architecture~\cite{ronneberger2015unet,rombach2021highresolution}, which comprises 8 channels and utilizes a cross-attention dimension of 1024.

\subsubsection{Trainable Module} The trainable modules include the control condition encoder and Fusion-Net. The control condition encoder incorporates a fixed text condition encoder that utilizes the weights from \textit{FLAN-T5-LARGE}~\cite{chung2022siflm}.  We employ 2-D convolutional layers with different kernel sizes and strides (2, 4, and 8) to transform the control condition tokens into different dimensions.

\subsubsection{Vocoder} To convert the generated mel-spectrogram into audio, we employ a vocoder. For fairness and consistency, we utilize HiFi-GAN~\cite{kong2020hifigan} that employs a set of small sub-discriminators to effectively address diverse cycle patterns as the vocoder following AudioLDM~\cite{liu2023audioldm} and Tango~\cite{ghosal2023tango}.

\subsubsection{Hyperparameters and Training Details} We refer to Tango~\cite{ghosal2023tango} and employ AdamW optimizer~\cite{loshchilov2017decoupledwd}. The learning rate is set to 3e-5, and a linear learning rate scheduler is utilized. Our model is trained for 30 epochs on 6 RTX3090 GPUs with the batch size of 2 per GPU, and we select the checkpoint with the lowest validation loss.

\subsection{B Visual Analysis}

\begin{figure}[!t]
    \centering
    \includegraphics[width=0.47\textwidth]{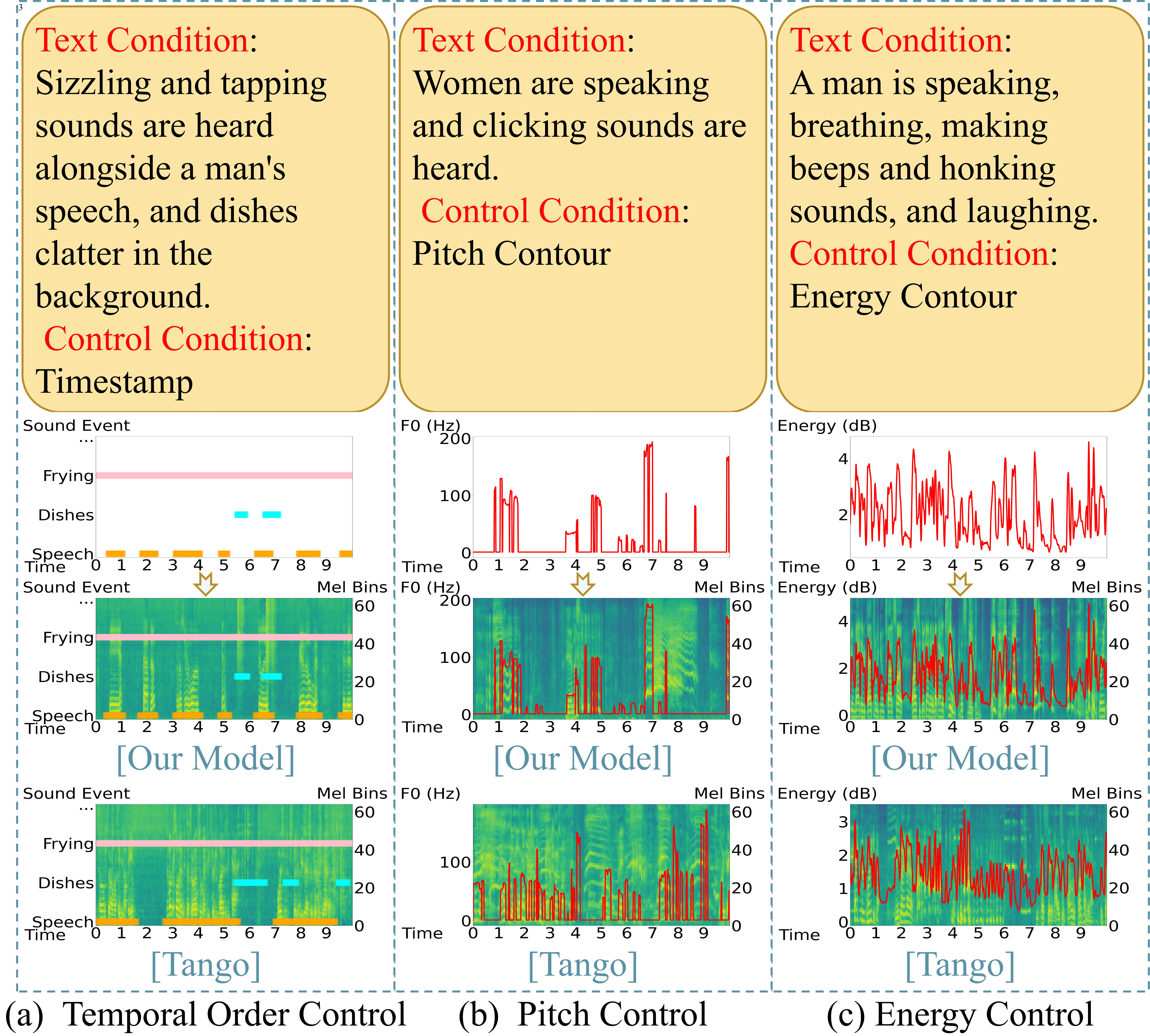}
    \caption{Comparison of mel-spectrograms generated by different models. Our model takes both text and control conditions as input, while Tango only takes the text condition as input. We show temporal order control in Subfigure (a), pitch control in Subfigure (b), and energy control in Subfigure (c).}
    \label{visual}
    \vspace{-4pt}
\end{figure}

In this section, we perform a visual analysis of the mel-spectrograms generated by our model and Tango~\cite{choi2018stargan}. Figure~\ref{visual} illustrates the comparison between the two models. While our model takes both text and control conditions as input, Tango exclusively utilizes text conditions as input. The mel-spectrograms produced by our model demonstrate a high level of conformity to the information provided in the control condition, while the mel-spectrograms generated by Tango exhibit noticeable deviations.

\bibliography{aaai24}

\end{document}